\documentclass[12pt]{article}
\usepackage{amsmath,amssymb, amsfonts, amsthm,amscd}
\usepackage[T2A]{fontenc}
\usepackage[cp1251]{inputenc}
\usepackage[russian,ukrainian,english]{babel}
\usepackage{geometry}

\newtheorem{Theorem}{Theorem}
\newtheorem{Corollary}{Corollary}

\newtheorem{Lemma}{Lemma}

\newenvironment{LemmaProof}{\textbf{Proof. }}{\par\noindent\textbf{The Lemma is proved.}}

\title{{\Large \textbf{On the $\mu$-parameters of the Petersen graph}}}

\author{\normalsize N.N. Davtyan}

\date{\small{\small{Ijevan Branch of Yerevan State University, e-mail:
nndavtyan@gmail.com}}}

\begin{document}

\maketitle

\bigskip

\begin{abstract}
For an undirected, simple, finite, connected graph $G$, we denote by
$V(G)$ and $E(G)$ the sets of its vertices and edges, respectively.
A function $\varphi:E(G)\rightarrow \{1,...,t\}$ is called a proper
edge $t$-coloring of a graph $G$, if adjacent edges are colored
differently and each of $t$ colors is used. The least value of $t$
for which there exists a proper edge $t$-coloring of a graph $G$ is
denoted by $\chi'(G)$. For any graph $G$, and for any integer $t$
satisfying the inequality $\chi'(G)\leq t\leq |E(G)|$, we denote by
$\alpha(G,t)$ the set of all proper edge $t$-colorings of $G$. Let
us also define a set $\alpha(G)$ of all proper edge colorings of a
graph $G$:
$$
\alpha(G)\equiv\bigcup_{t=\chi'(G)}^{|E(G)|}\alpha(G,t).
$$

An arbitrary nonempty finite subset of consecutive integers is
called an interval. If $\varphi\in\alpha(G)$ and $x\in V(G)$, then
the set of colors of edges of $G$ which are incident with $x$ is
denoted by $S_G(x,\varphi)$ and is called a spectrum of the vertex
$x$ of the graph $G$ at the proper edge coloring $\varphi$. If $G$
is a graph and $\varphi\in\alpha(G)$, then define
$f_G(\varphi)\equiv|\{x\in V(G)/S_G(x,\varphi) \textrm{ is an
interval}\}|$.

For a graph $G$ and any integer $t$, satisfying the inequality
$\chi'(G)\leq t\leq |E(G)|$, we define:
$$
\mu_1(G,t)\equiv\min_{\varphi\in\alpha(G,t)}f_G(\varphi),\qquad
\mu_2(G,t)\equiv\max_{\varphi\in\alpha(G,t)}f_G(\varphi).
$$

For any graph $G$, we set:
$$
\mu_{11}(G)\equiv\min_{\chi'(G)\leq t\leq|E(G)|}\mu_1(G,t),\qquad
\mu_{12}(G)\equiv\max_{\chi'(G)\leq t\leq|E(G)|}\mu_1(G,t),
$$
$$
\mu_{21}(G)\equiv\min_{\chi'(G)\leq t\leq|E(G)|}\mu_2(G,t),\qquad
\mu_{22}(G)\equiv\max_{\chi'(G)\leq t\leq|E(G)|}\mu_2(G,t).
$$

For the Petersen graph, the exact values of the parameters
$\mu_{11}$, $\mu_{12}$, $\mu_{21}$ and $\mu_{22}$ are found.

\bigskip
Keywords: the Petersen graph, proper edge coloring, interval
spectrum, game.

Math. Classification: 05C15
\end{abstract}

We consider finite, undirected, connected graphs without loops and
multiple edges containing at least one edge. For any graph $G$, we
denote by $V(G)$ and $E(G)$ the sets of vertices and edges of $G$,
respectively. For any $x\in V(G)$, $d_G(x)$ denotes the degree of
the vertex $x$ in $G$. For a graph $G$, $\delta(G)$ and $\Delta(G)$
denote the minimum and maximum degrees of vertices in $G$,
respectively. For a graph $G$, and for any $V_0\subseteq V(G)$, we
denote by $G[V_0]$ the subgraph of the graph $G$ induced
\cite{West1} by the subset $V_0$ of its vertices.

An arbitrary nonempty finite subset of consecutive integers is
called an interval. An interval with the minimum element $p$ and the
maximum element $q$ is denoted by $[p,q]$.

A function $\varphi:E(G)\rightarrow [1,t]$ is called a proper edge
$t$-coloring of a graph $G$, if each of $t$ colors is used, and
adjacent edges are colored differently.

The minimum value of $t$ for which there exists a proper edge
$t$-coloring of a graph $G$ is denoted by $\chi'(G)$ \cite{Vizing2}.

We denote by $P$ \cite{BookP} the Petersen graph. $P$ is a cubic
graph with $|V(P)|=10$, $|E(P)|=15$, $\chi'(P)=\Delta(P)+1=4$. In
this paper we assume that
$$
\begin{array}{l}
V(P)=\{x_1,x_2,x_3,x_4,x_5,y_1,y_2,y_3,y_4,y_5\},\\
E(P)=\{(x_1,x_2),(x_2,x_3),(x_3,x_4),(x_4,x_5),(x_1,x_5),(x_1,y_1),(x_2,y_2),
(x_3,y_3),(x_4,y_4),(x_5,y_5),\\
(y_1,y_3),(y_1,y_4),(y_2,y_4),(y_2,y_5),(y_3,y_5)\}.
\end{array}
$$

For any graph $G$, and for any $t\in[\chi'(G),|E(G)|]$, we denote by
$\alpha(G,t)$ the set of all proper edge $t$-colorings of $G$.

Let us also define a set $\alpha(G)$ of all proper edge colorings of
a graph $G$:
$$
\alpha(G)\equiv\bigcup_{t=\chi'(G)}^{|E(G)|}\alpha(G,t).
$$

If $\varphi\in\alpha(G)$ and $x\in V(G)$, then the set
$\{\varphi(e)/ e\in E(G), e \textrm{ is incident with } x$\} is
called a spectrum of the vertex $x$ of the graph $G$ at the proper
edge coloring $\varphi$ and is denoted by $S_G(x,\varphi)$.

If $G$ is a graph, $\varphi\in\alpha(G)$, then set
$V_{int}(G,\varphi)\equiv\{x\in V(G)/S_G(x,\varphi) \textrm{ is an
interval}\}$ and $f_G(\varphi)\equiv|V_{int}(G,\varphi)|$. A proper
edge coloring $\varphi\in\alpha(G)$ is called an interval edge
coloring \cite{Oranj3, Asratian4, Diss5} of the graph $G$ iff
$f_G(\varphi)=|V(G)|$. The set of all graphs having an interval edge
coloring is denoted by $\mathfrak{N}$. The simplest example of the
graph which doesn't belong to $\mathfrak{N}$ is $K_3$. The terms and
concepts which are not defined can be found in \cite{West1}.

For a graph $G$, and for any $t\in[\chi'(G),|E(G)|]$, we set
\cite{Mebius6}:
$$
\mu_1(G,t)\equiv\min_{\varphi\in\alpha(G,t)}f_G(\varphi),\qquad
\mu_2(G,t)\equiv\max_{\varphi\in\alpha(G,t)}f_G(\varphi).
$$

For any graph $G$, we set \cite{Mebius6}:
$$
\mu_{11}(G)\equiv\min_{\chi'(G)\leq t\leq|E(G)|}\mu_1(G,t),\qquad
\mu_{12}(G)\equiv\max_{\chi'(G)\leq t\leq|E(G)|}\mu_1(G,t),
$$
$$
\mu_{21}(G)\equiv\min_{\chi'(G)\leq t\leq|E(G)|}\mu_2(G,t),\qquad
\mu_{22}(G)\equiv\max_{\chi'(G)\leq t\leq|E(G)|}\mu_2(G,t).
$$

Clearly, the parameters $\mu_{11}$, $\mu_{12}$, $\mu_{21}$ and
$\mu_{22}$ are correctly defined for an arbitrary graph.

Let us note that exact values of the parameters $\mu_{12}$ and
$\mu_{21}$ have certain game interpretations. Suppose that all edges
of a graph $G$ are colored in the game of Alice and Bob with
antagonistic interests and asymmetric distribution of roles. Alice
determines the number $t$ of colors in the future proper edge
coloring $\varphi$ of the graph $G$, satisfying the condition
$t\in[\chi'(G),|E(G)|]$, Bob colors edges of $G$ with $t$ colors.

When Alice aspires to maximize, Bob aspires to minimize the value of
the function $f_G(\varphi)$, and both players choose their best
strategies, then at the finish of the game exactly $\mu_{12}(G)$
vertices of $G$ will receive an interval spectrum.

When Alice aspires to minimize, Bob aspires to maximize the value of
the function $f_G(\varphi)$, and both players choose their best
strategies, then at the finish of the game exactly $\mu_{21}(G)$
vertices of $G$ will receive an interval spectrum.

The exact values of the parameters $\mu_{11}$, $\mu_{12}$,
$\mu_{21}$ and $\mu_{22}$ are found for simple paths, simple cycles
and simple cycles with a chord \cite{Simple7, Akunq}, "M\"{o}bius
ladders" \cite{Mebius6, Minchev}, complete graphs \cite{Arpine8},
complete bipartite graphs \cite{Arpine9, Arpine10}, prisms
\cite{Arpine11, Minchev} and $n$-dimensional cubes \cite{Arpine11,
Nikolaev12, KornArxive}. The exact values of $\mu_{11}$ and
$\mu_{22}$ for trees are found in \cite{Evg13}. The exact value of
$\mu_{12}$ for an arbitrary tree is found in \cite{Trees14} (see
also \cite{Algorithm, Tree_Kontr}).

In this paper we determine the exact values of the parameters
$\mu_{11}$, $\mu_{12}$, $\mu_{21}$ and $\mu_{22}$ for the Petersen
graph $P$.

First we recall some known results.

\begin{Lemma}\cite{Oranj3, Asratian4, Diss5}\label{lem1}
If $G\in\mathfrak{N}$, then $\chi'(G)=\Delta(G)$.
\end{Lemma}

\begin{Corollary}\label{cor1}
If $G$ is a regular graph, then $G\in\mathfrak{N}$ iff
$\chi'(G)=\Delta(G)$.
\end{Corollary}

\begin{Corollary}\label{cor1'}
$P\not\in\mathfrak{N}$.
\end{Corollary}

\begin{Corollary}\label{cor1''}
$\mu_{22}(P)\leq9$.
\end{Corollary}

\begin{Lemma}\cite{Luhansk17}\label{lem2}
If $G$ is a graph with $\delta(G)\geq 2$,
$\varphi\in\alpha(G,|E(G)|)$, $V_{int}(G,\varphi)\neq\emptyset$,
then $G[V_{int}(G,\varphi)]$ is a forest each connected component of
which is a simple path.
\end{Lemma}

\begin{Lemma}\label{lem3}
If $F_1$ and $F_2$ are two arbitrary perfect matchings of $P$, then
$F_1\cap F_2\neq\emptyset$.
\end{Lemma}

\begin{Lemma}\label{lem4}
If a subset $V_0$ of the set of vertices of the graph $P$ contains
at least $7$ vertices, then at least one of the following two
statements is true:
\begin{enumerate}
  \item there exist such vertices $a_1, a_2, a_3, a_4$ in $V_0$ that $P[\{a_1, a_2, a_3, a_4\}]\cong K_{3,1}$,
  \item there exist such vertices $b_1, b_2, b_3, b_4, b_5, b_6$ in $V_0$ that $P[\{b_1, b_2, b_3, b_4, b_5, b_6\}]\cong C_6$.
\end{enumerate}
\end{Lemma}

\textbf{Proof} is evident.

\begin{Lemma}\label{lem5}
There exists $\varphi\in\alpha(P,15)$ with $f_P(\varphi)=0$. There
exists $\psi\in\alpha(P,15)$ with $f_P(\psi)=6$. There exists
$\varepsilon\in\alpha(P,4)$ with $f_P(\varepsilon)=2$. There exists
$\sigma\in\alpha(P,4)$ with $f_P(\sigma)=8$.
\end{Lemma}

\begin{LemmaProof}

Set:
$$
\varphi((x_1,x_2))=1,\quad \varphi((x_1,y_1))=2, \quad
\varphi((y_1,y_3))=3, \quad \varphi((x_1,x_5))=4, \quad
\varphi((x_5,y_5))=5,
$$
$$
\varphi((y_1,y_4))=6, \quad \varphi((x_4,x_5))=7, \quad
\varphi((x_4,y_4))=8, \quad \varphi((y_2,y_5))=9, \quad
\varphi((x_3,x_4))=10,
$$
$$
\varphi((x_3,y_3))=11, \quad \varphi((y_3,y_5))=12, \quad
\varphi((x_2,x_3))=13, \quad \varphi((x_2,y_2))=14, \quad
\varphi((y_2,y_4))=15.
$$

It is not difficult to see that $\varphi\in\alpha(P,15)$ and
$f_P(\varphi)=0$.

\bigskip

Set:
$$
\psi((y_1,y_3))=1,\quad \psi((y_3,y_5))=2, \quad \psi((x_3,y_3))=3,
\quad \psi((x_2,x_3))=4, \quad \psi((x_3,x_4))=5,
$$
$$
\psi((x_4,y_4))=6, \quad \psi((x_4,x_5))=7, \quad \psi((x_5,y_5))=8,
\quad \psi((x_1,x_5))=9, \quad \psi((x_1,y_1))=10,
$$
$$
\psi((x_1,x_2))=11, \quad \psi((x_2,y_2))=12, \quad
\psi((y_2,y_5))=13, \quad \psi((y_2,y_4))=14, \quad
\psi((y_1,y_4))=15.
$$

It is not difficult to see that $\psi\in\alpha(P,15)$ and
$f_P(\psi)=6$.

\bigskip

Set:
$$
\varepsilon((x_1,y_1))=\varepsilon((x_2,x_3))=\varepsilon((y_3,y_5))=\varepsilon((x_4,x_5))=\varepsilon((y_2,y_4))=1,
$$
$$
\varepsilon((x_1,x_2))=\varepsilon((x_3,x_4))=\varepsilon((y_2,y_5))=2,
$$
$$
\varepsilon((y_1,y_4))=\varepsilon((x_3,y_3))=\varepsilon((x_5,y_5))=3,
$$
$$
\varepsilon((x_1,x_5))=\varepsilon((y_1,y_3))=\varepsilon((x_4,y_4))=\varepsilon((x_2,y_2))=4.
$$

It is not difficult to see that $\varepsilon\in\alpha(P,4)$ and
$f_P(\varepsilon)=2$.

\bigskip

Set:
$$
\sigma((y_1,y_4))=\sigma((y_3,y_5))=1,
$$
$$
\sigma((x_1,x_2))=\sigma((y_1,y_3))=\sigma((x_3,x_4))=\sigma((y_2,y_4))=\sigma((x_5,y_5))=2,
$$
$$
\sigma((x_2,y_2))=\sigma((x_3,y_3))=\sigma((x_4,y_4))=\sigma((x_1,x_5))=3,
$$
$$
\sigma((x_1,y_1))=\sigma((x_2,x_3))=\sigma((x_4,x_5))=\sigma((y_2,y_5))=4.
$$

It is not difficult to see that $\sigma\in\alpha(P,4)$ and
$f_P(\sigma)=8$.

\end{LemmaProof}

\begin{Corollary}\label{cor2}
$\mu_1(P,15)=0$, $\mu_2(P,15)\geq6$, $\mu_1(P,4)\leq2$,
$\mu_2(P,4)\geq8$.
\end{Corollary}

\begin{Corollary}\label{cor3}
$\mu_{11}(P)=0$, $\mu_{22}(P)\geq8$.
\end{Corollary}

\begin{Corollary}\label{cor3'}
$8\leq\mu_{22}(P)\leq9$.
\end{Corollary}

\begin{Lemma}\label{lem6}
$\mu_2(P,15)\leq6$.
\end{Lemma}

\begin{LemmaProof}
Assume the contrary. Then there exists $\varphi_0\in\alpha(P,15)$,
for which $f_P(\varphi_0)\geq7$.

By lemma \ref{lem2}, $P[V_{int}(P,\varphi_0)]$ is a forest, each
connected component of which is a simple path. But it is
incompatible with lemma \ref{lem4}.
\end{LemmaProof}

From corollary \ref{cor2} and lemma \ref{lem6} we obtain

\begin{Lemma}\label{lem7}
$\mu_2(P,15)=6$.
\end{Lemma}

\begin{Corollary}\label{cor4}
$\mu_{21}(P)\leq6$.
\end{Corollary}

\begin{Lemma}\label{lem8}
$\mu_2(P,14)\geq6$. $\mu_2(P,13)\geq6$. $\mu_2(P,12)\geq6$.
$\mu_2(P,11)\geq6$. $\mu_2(P,10)\geq6$. $\mu_2(P,9)\geq6$.
$\mu_2(P,8)\geq6$. $\mu_2(P,7)\geq7$. $\mu_2(P,6)\geq7$.
$\mu_2(P,5)\geq7$. $\mu_2(P,4)\geq8$.
\end{Lemma}

\begin{LemmaProof}
Let us construct the sequence of proper edge colorings $\psi_0$,
$\psi_1$, $\psi_2$, $\psi_3$, $\psi_4$, $\psi_5$, $\psi_6$,
$\psi_7$, $\psi_8$, $\psi_9$, $\psi_{10}$, $\psi_{11}$ of $P$
defined as follows.

$\psi_0\equiv\psi$, where $\psi$ is the proper edge $15$-coloring
constructed for the proof of lemma \ref{lem5}.

Let us define $\psi_1$.

For $\forall e\in E(P)$, set:

$$
\psi_1(e)\equiv\left\{
\begin{array}{ll}
2, & \textrm{if $\;e=(y_1,y_4)$},\\
\psi_0(e) & \textrm{-- otherwise}.\\
\end{array}
\right.
$$

Clearly, $\psi_1\in\alpha(P,14)$ and $f_P(\psi_1)=6$. Consequently,
$\mu_2(P,14)\geq6$.

\bigskip

Let us define $\psi_2$.

For $\forall e\in E(P)$, set:

$$
\psi_2(e)\equiv\left\{
\begin{array}{ll}
11, & \textrm{if $\;e=(y_2,y_4)$},\\
\psi_1(e) & \textrm{-- otherwise}.\\
\end{array}
\right.
$$

Clearly, $\psi_2\in\alpha(P,13)$ and $f_P(\psi_2)=6$. Consequently,
$\mu_2(P,13)\geq6$.

\bigskip

Let us define $\psi_3$.

For $\forall e\in E(P)$, set:

$$
\psi_3(e)\equiv\left\{
\begin{array}{ll}
10, & \textrm{if $\;e=(y_2,y_5)$},\\
\psi_2(e) & \textrm{-- otherwise}.\\
\end{array}
\right.
$$

Clearly, $\psi_3\in\alpha(P,12)$ and $f_P(\psi_3)=6$. Consequently,
$\mu_2(P,12)\geq6$.

\bigskip

Let us define $\psi_4$.

For $\forall e\in E(P)$, set:

$$
\psi_4(e)\equiv\left\{
\begin{array}{ll}
9, & \textrm{if $\;e=(x_2,y_2)$},\\
\psi_3(e) & \textrm{-- otherwise}.\\
\end{array}
\right.
$$

Clearly, $\psi_4\in\alpha(P,11)$ and $f_P(\psi_4)=6$. Consequently,
$\mu_2(P,11)\geq6$.

\bigskip

Let us define $\psi_5$.

For $\forall e\in E(P)$, set:

$$
\psi_5(e)\equiv\left\{
\begin{array}{ll}
8, & \textrm{if $\;e=(x_1,x_2)$ or $\;e=(y_2,y_4)$},\\
\psi_4(e) & \textrm{-- otherwise}.\\
\end{array}
\right.
$$

Clearly, $\psi_5\in\alpha(P,10)$ and $f_P(\psi_5)=6$. Consequently,
$\mu_2(P,10)\geq6$.

\bigskip

Let us define $\psi_6$.

For $\forall e\in E(P)$, set:

$$
\psi_6(e)\equiv\left\{
\begin{array}{ll}
7, & \textrm{if $\;e=(x_1,y_1)$ or $\;e=(y_2,y_5)$},\\
\psi_5(e) & \textrm{-- otherwise}.\\
\end{array}
\right.
$$

Clearly, $\psi_6\in\alpha(P,9)$ and $f_P(\psi_6)=6$. Consequently,
$\mu_2(P,9)\geq6$.

\bigskip

Let us define $\psi_7$.

For $\forall e\in E(P)$, set:

$$
\psi_7(e)\equiv\left\{
\begin{array}{ll}
6, & \textrm{if $\;e=(x_1,x_5)$ or $\;e=(x_2,y_2)$},\\
\psi_6(e) & \textrm{-- otherwise}.\\
\end{array}
\right.
$$

Clearly, $\psi_7\in\alpha(P,8)$ and $f_P(\psi_7)=6$. Consequently,
$\mu_2(P,8)\geq6$.

\bigskip

Let us define $\psi_8$.

For $\forall e\in E(P)$, set:

$$
\psi_8(e)\equiv\left\{
\begin{array}{ll}
5, & \textrm{if $\;e=(x_1,x_2)$, $\;e=(x_5,y_5)$ or $\;e=(y_2,y_4)$},\\
\psi_7(e) & \textrm{-- otherwise}.\\
\end{array}
\right.
$$

Clearly, $\psi_8\in\alpha(P,7)$ and $f_P(\psi_8)=7$. Consequently,
$\mu_2(P,7)\geq7$.

\bigskip

Let us define $\psi_9$.

For $\forall e\in E(P)$, set:

$$
\psi_9(e)\equiv\left\{
\begin{array}{ll}
4, & \textrm{if $\;e=(x_1,y_1)$, $\;e=(x_4,x_5)$ or $\;e=(y_2,y_5)$},\\
\psi_8(e) & \textrm{-- otherwise}.\\
\end{array}
\right.
$$

Clearly, $\psi_9\in\alpha(P,6)$ and $f_P(\psi_9)=7$. Consequently,
$\mu_2(P,6)\geq7$.

\bigskip

Let us define $\psi_{10}$.

For $\forall e\in E(P)$, set:

$$
\psi_{10}(e)\equiv\left\{
\begin{array}{ll}
3, & \textrm{if $\;e=(x_1,x_5)$, $\;e=(x_4,y_4)$ or $\;e=(x_2,y_2)$},\\
\psi_9(e) & \textrm{-- otherwise}.\\
\end{array}
\right.
$$

Clearly, $\psi_{10}\in\alpha(P,5)$ and $f_P(\psi_{10})=7$.
Consequently, $\mu_2(P,5)\geq7$.

\bigskip

$\psi_{11}\equiv\sigma$, where $\sigma$ is the proper edge
$4$-coloring constructed for the proof of lemma \ref{lem5}.
\end{LemmaProof}

From lemmas \ref{lem7} and \ref{lem8} we obtain

\begin{Lemma}\label{lem9}
$\mu_{21}(P)=6$.
\end{Lemma}

From \cite{Steffen} we have

\begin{Lemma}\label{lem11}
An arbitrary graph $H$ obtained from the graph $P$ by removing of
its one vertex, satisfies the condition $\chi'(H)=4$.
\end{Lemma}

\begin{Lemma}\label{lem12}
$\mu_{22}(P)=8$.
\end{Lemma}

\begin{LemmaProof}
Assume the contrary. Then, by corollary \ref{cor3'}, we have
$\mu_{22}(P)=9$. It means that there exists $t_0\in[4,14]$, for
which $\mu_2(P,t_0)=9$. Consequently, there exists
$\widetilde{\varphi}\in\alpha(P,t_0)$ with
$f_P(\widetilde{\varphi})=9$. Let us define the subsets
$E_1,E_2,E_3$ of the set $E(P)$ as follows:
$$
\begin{array}{l}
E_1\equiv\{e\in E(P)/\;\widetilde{\varphi}\equiv1(mod3)\},\\
E_2\equiv\{e\in E(P)/\;\widetilde{\varphi}\equiv2(mod3)\},\\
E_3\equiv\{e\in E(P)/\;\widetilde{\varphi}\equiv0(mod3)\}.
\end{array}
$$

Clearly, $E_1\cup E_2\cup E_3=E(P)$, $E_1\cap E_2=\emptyset$,
$E_1\cap E_3=\emptyset$, $E_2\cap E_3=\emptyset$.

Let $H\equiv P[V_{int}(P,\widetilde{\varphi})]$. Clearly,
$|V(H)|=9$, $|E(H)|=12$, $\Delta(H)=3$, $\delta(H)=2$. Evidently,
$H$ can be obtained from $P$ by removing of its one vertex.

Let us define a function $\xi : E(H)\rightarrow[1,3]$ as follows.
For $\forall e\in E(H)$, set:
$$
\xi(e)\equiv\left\{
\begin{array}{ll}
1, & \textrm{if $\;e\in E_1$},\\
2, & \textrm{if $\;e\in E_2$},\\
3, & \textrm{if $\;e\in E_3$}.\\
\end{array}
\right.
$$

It is not difficult to see that $\xi\in\alpha(H,3)$, and,
consequently, $\chi'(H)=3$. It contradicts lemma \ref{lem11}.
\end{LemmaProof}

\begin{Lemma}\label{lem13}
$\mu_1(P,4)\geq2$.
\end{Lemma}

\begin{LemmaProof}
Assume the contrary: $\mu_1(P,4)\leq1$. It means that there exists
$\beta\in\alpha(P,4)$ with $f_P(\beta)\leq1$. It implies the
inequality $|\{z\in V(P)/\;\{1,4\}\subset S_P(z,\beta)\}|\geq9$. It
means that the subsets $\{e\in E(P)/\;\beta(e)=1\}$ and $\{e\in
E(P)/\;\beta(e)=4\}$ of edges of $P$ are both perfect matchings in
$P$. It contradicts lemma \ref{lem3}.
\end{LemmaProof}

From corollary \ref{cor2} and lemma \ref{lem13} we obtain

\begin{Corollary}\label{cor6}
$\mu_1(P,4)=2$.
\end{Corollary}

\begin{Lemma}\label{lem14}
For $\forall t\in[5,14]$, $\mu_1(P,t)=0$.
\end{Lemma}

\begin{LemmaProof}
Let us construct the sequence of proper edge colorings $\lambda_0$,
$\lambda_1$, $\lambda_2$, $\lambda_3$, $\lambda_4$, $\lambda_5$,
$\lambda_6$, $\lambda_7$, $\lambda_8$, $\lambda_9$, $\lambda_{10}$
of $P$ defined as follows.

$\lambda_0\equiv\varepsilon$, where $\varepsilon$ is the proper edge
$4$-coloring constructed for the proof of lemma \ref{lem5}.

Let us define $\lambda_1$.

For $\forall e\in E(P)$, set:

$$
\lambda_1(e)\equiv\left\{
\begin{array}{ll}
5, & \textrm{if $\;e=(x_2,x_3)$ or $\;e=(y_2,y_5)$},\\
\lambda_0(e) & \textrm{-- otherwise}.\\
\end{array}
\right.
$$

Clearly, $\lambda_1\in\alpha(P,5)$ and $f_P(\lambda_1)=0$.
Consequently, $\mu_1(P,5)=0$.

\bigskip

Let us define $\lambda_2$.

For $\forall e\in E(P)$, set:

$$
\lambda_2(e)\equiv\left\{
\begin{array}{ll}
6, & \textrm{if $\;e=(y_3,y_5)$},\\
\lambda_1(e) & \textrm{-- otherwise}.\\
\end{array}
\right.
$$

Clearly, $\lambda_2\in\alpha(P,6)$ and $f_P(\lambda_2)=0$.
Consequently, $\mu_1(P,6)=0$.

\bigskip

Let us define $\lambda_3$.

For $\forall e\in E(P)$, set:

$$
\lambda_3(e)\equiv\left\{
\begin{array}{ll}
7, & \textrm{if $\;e=(y_2,y_4)$},\\
\lambda_2(e) & \textrm{-- otherwise}.\\
\end{array}
\right.
$$

Clearly, $\lambda_3\in\alpha(P,7)$ and $f_P(\lambda_3)=0$.
Consequently, $\mu_1(P,7)=0$.

\bigskip

Let us define $\lambda_4$.

For $\forall e\in E(P)$, set:

$$
\lambda_4(e)\equiv\left\{
\begin{array}{ll}
8, & \textrm{if $\;e=(x_4,x_5)$},\\
\lambda_3(e) & \textrm{-- otherwise}.\\
\end{array}
\right.
$$

Clearly, $\lambda_4\in\alpha(P,8)$ and $f_P(\lambda_4)=0$.
Consequently, $\mu_1(P,8)=0$.

\bigskip

Let us define $\lambda_5$.

For $\forall e\in E(P)$, set:

$$
\lambda_5(e)\equiv\left\{
\begin{array}{ll}
9, & \textrm{if $\;e=(x_1,x_2)$},\\
\lambda_4(e) & \textrm{-- otherwise}.\\
\end{array}
\right.
$$

Clearly, $\lambda_5\in\alpha(P,9)$ and $f_P(\lambda_5)=0$.
Consequently, $\mu_1(P,9)=0$.

\bigskip

Let us define $\lambda_6$.

For $\forall e\in E(P)$, set:

$$
\lambda_6(e)\equiv\left\{
\begin{array}{ll}
10, & \textrm{if $\;e=(x_5,y_5)$},\\
\lambda_5(e) & \textrm{-- otherwise}.\\
\end{array}
\right.
$$

Clearly, $\lambda_6\in\alpha(P,10)$ and $f_P(\lambda_6)=0$.
Consequently, $\mu_1(P,10)=0$.

\bigskip

Let us define $\lambda_7$.

For $\forall e\in E(P)$, set:

$$
\lambda_7(e)\equiv\left\{
\begin{array}{ll}
11, & \textrm{if $\;e=(x_3,y_3)$},\\
\lambda_6(e) & \textrm{-- otherwise}.\\
\end{array}
\right.
$$

Clearly, $\lambda_7\in\alpha(P,11)$ and $f_P(\lambda_7)=0$.
Consequently, $\mu_1(P,11)=0$.

\bigskip

Let us define $\lambda_8$.

For $\forall e\in E(P)$, set:

$$
\lambda_8(e)\equiv\left\{
\begin{array}{ll}
12, & \textrm{if $\;e=(x_4,y_4)$},\\
\lambda_7(e) & \textrm{-- otherwise}.\\
\end{array}
\right.
$$

Clearly, $\lambda_8\in\alpha(P,12)$ and $f_P(\lambda_8)=0$.
Consequently, $\mu_1(P,12)=0$.

\bigskip

Let us define $\lambda_9$.

For $\forall e\in E(P)$, set:

$$
\lambda_9(e)\equiv\left\{
\begin{array}{ll}
13, & \textrm{if $\;e=(y_1,y_3)$},\\
\lambda_8(e) & \textrm{-- otherwise}.\\
\end{array}
\right.
$$

Clearly, $\lambda_9\in\alpha(P,13)$ and $f_P(\lambda_9)=0$.
Consequently, $\mu_1(P,13)=0$.

\bigskip

Let us define $\lambda_{10}$.

For $\forall e\in E(P)$, set:

$$
\lambda_{10}(e)\equiv\left\{
\begin{array}{ll}
14, & \textrm{if $\;e=(x_2,y_2)$},\\
\lambda_9(e) & \textrm{-- otherwise}.\\
\end{array}
\right.
$$

Clearly, $\lambda_{10}\in\alpha(P,14)$ and $f_P(\lambda_{10})=0$.
Consequently, $\mu_1(P,14)=0$.
\end{LemmaProof}

From lemmas \ref{lem5} and \ref{lem14}, corollary \ref{cor6} we
obtain

\begin{Corollary}\label{cor5}
$\mu_{12}(P)=2$.
\end{Corollary}

From corollaries \ref{cor3}, \ref{cor5} and lemmas \ref{lem9},
\ref{lem12} we obtain

\begin{Theorem}
For the Petersen graph $P$, the equalities $\mu_{11}(P)=0$,
$\mu_{12}(P)=2$, $\mu_{21}(P)=6$ and $\mu_{22}(P)=8$ are true.
\end{Theorem}

\bigskip\bigskip

\textbf{Acknowledgement.} I thank my scientific supervisor R.R.
Kamalian for his constant moral support and useful advices.

\end{document}